\documentclass[prb,twocolumn,showpacs,preprintnumbers,amsmath,amssymb]{revtex4}
\usepackage{amssymb}
\usepackage{graphicx}
\usepackage{amsmath}

\begin{document}
 
\title{Role of the transverse field in inverse freezing in the fermionic
    Ising spin-glass model}
 
\author{S. G. Magalhaes}\email{ggarcia@ccne.ufsm.br} \author{C. V. Morais}  \affiliation{
Universidade Federal de Santa Maria, Santa Maria, Rio Grande do Sul 97105-900, Brazil}%
\author{F. M. Zimmer}
 \affiliation{Departamento de F\'isica, Universidade do Estado de Santa Catarina, Joinville, Santa Catarina 89223-100, Brazil
 }

\begin{abstract}
We investigate the inverse freezing in the fermionic Ising spin-glass (FISG) model in a transverse field $\Gamma$. The grand canonical potential is calculated in the static approximation, replica symmetry and one-step replica symmetry 
 breaking Parisi scheme. It is argued that the average occupation  per site $n$ is 
strongly
 affected by $\Gamma$. 
As consequence,
the boundary phase is modified and, 
therefore,
the reentrance associated  with the inverse freezing  is modified too.
\end{abstract}
\maketitle

The 
inverse transitions (melting or freezing), first proposed by Tammann \cite{Tammann},
are
a class of a quite interesting phase transitions,
utterly counterintuitive,
in which the ordered phase has more 
entropy than the disordered one \cite{Schupper1}. 
Despite this apparent 
unconventional thermodynamics,
there is now a list plenty of physical systems in which this sort of transition  appears (see Ref. \onlinecite{Schupper1} and references therein)
 including, interestingly, high temperature superconductors.
\cite{Avraham} In that sense, 
the search for theoretical models which contain the necessary 
ingredients to produce such transitions 
 has become a 
challenging issue
as can be seen in Refs. \onlinecite{Schupper1} and \onlinecite{Crisanti,Sellito,Prestipino}.
However, 
much less
considerations have been given to models in which quantum effects can also be taken into account.
 
It should be noticed that
there is an important difference among some of the previous mentioned models.
For example, 
in Refs. \onlinecite{Schupper1}, Schupper and Shnerb show that the Blume-Capel model
can present inverse melting
since 
the entropic advantage of the interacting state  is introduced in the problem 
 by imposing that the degeneracy parameter $r=k/l\geq 1$, where
it is assumed that 
 $\pm 1$ spin states are $k$-fold degenerated, 
while $0$ spin states are $l$-fold degenerated.
In 
the 
Refs. \onlinecite{Schupper1} and \onlinecite{Crisanti}, the classical Ghatak-Sherrington \cite{GS} (GS) model has also been proposed to be one of the simplest disordered models to present inverse freezing. 
Nevertheless, for the GS model,
as remarked in Ref. \onlinecite{Crisanti}, 
there is no need to enforce the entropic advantage.
Thus, the first order boundary of the spin-glass (SG) and/or paramagnetic (PM) transition
naturally
displays a 
reentrance.
That means 
it is possible to enter in the SG 
phase
by heating from 
the
PM one.
That raises 
the following questions: 
is it possible to find 
other
disordered models with
inverse freezing 
but
no additional enforcement of 
the entropic advantage?  
Would it also  be possible to 
incorporate
quantum effects? 

It is now well known that  
the classical GS model 
has a very close relationship with the fermionic Ising spin-glass (FISG)\cite{Oppermannfisg,Castillo} model
in
the static approximation (SA) \cite{BrayMoore}.
In the FISG model \cite{Albamercedes,Sherrington,Oppermannnuclear}, 
the spin operators are written by bilinear combination of fermionic creation and destruction operators which act on a space with four eigenstates per site $(|00\rangle,|\uparrow 0\rangle, |0\downarrow\rangle,|\uparrow\downarrow\rangle)$. 
In particular,
the thermodynamics of 
both models can be exactly mapped by a relationship between the anisotropic 
constant $D$  
and the chemical potential $\mu$ of GS and FISG models \cite{Oppermannfisg}, respectively.  
 Thus,  
we can expect strong resemblances 
between
the phase diagrams of the two models; particularly,  
the presence of  
reentrance in the first order boundary must be emphasized
(see, 
for instance, 
Ref. 
\onlinecite{Oppermannprb}). Therefore, the FISG model
can
 also be
considered
one
of the models which 
naturally presents 
inverse freezing. 
Actually, 
the FISG model 
has been intensively used
to study 
the competition between the SG phase 
and, for example, superconductivity or Kondo effect (see Refs. \onlinecite{OppermannSuper} and \onlinecite{MagalCoqblin} and references therein). 
Moreover, the FISG has also been used \cite{Albaphysicaa} to 
study  the effects of the   
transverse field $\Gamma$ on the PM/SG boundary phase  within SA.
At the half-filling ($\mu=0$), 
it has been shown that the increase of $\Gamma$ strongly
changes
the FISG thermodynamics since it 
tends
to suppress the SG phase leading 
the freezing temperature $T_{f}$ to a quantum critical point (QCP) at $\Gamma_{c}$. 
Therefore, the FISG model, besides allowing to treat charge and spin at  the same level, 
can be  a
useful tool to study 
inverse freezing,  
particularly,
when quantum effects can be included.

The goal 
of the present paper
is to investigate 
the inverse freezing in the FISG 
model when 
spin flipping is induced by a transverse field $\Gamma$.
It should be remarked that
a set of FISG 
order parameters 
is composed not only by the usual nondiagonal
SG order parameter $q_{\alpha\beta}$ ($\alpha\neq\beta$), 
but also by
the 
diagonal one $q_{\alpha\alpha}$, which is directly 
related 
to
the 
average occupation of fermions per site $n$
\cite{Oppermannnuclear}.
Since $\Gamma$
affects the behavior of the SG order parameters, 
we can assume
that $\Gamma$ could have strong influence on $n$ as well.
Actually,
the behavior of $n$ 
is determined by a more complicated dependence on $\Gamma$ than the obvious one 
given by $q_{\alpha\alpha}$.
In that view, the 
spin flipping produced by
$\Gamma$ can also be considered as a 
mechanism to change 
the 
charge  
occupation 
and, hence, 
to modify
the 
original 
phase boundary (when $\Gamma=0$) 
in the 
$T$-$\mu$ plane. 
In this sense, 
we
can 
probe 
the   
robustness 
of the 
inverse freezing 
when quantum effects are present, using what would be, in principle, a controlled external field.
In 
this work, 
the partition function is obtained within the Grassmann functional 
integral formalism \cite{Sherrington}. 
The SG order parameters are 
calculated within the SA, in the 
 replica symmetry (RS) and one-step replica symmetry 
breaking (1S-RSB) Parisi scheme.
Previous calculations have shown \cite{Crisanti} that the position of the first order reentrance 
has no significant difference when obtained in
RS, 1S-RSB 
or  even in the
full replica symmetry breaking (FRSB) Parisi scheme 
\cite{Parisi}. However, 
to be sure that the first order boundary 
is not affected by the RS instability when $\Gamma$ is present, we also 
obtain the thermodynamics in the 1S-RSB. 
It should also be
highlighted that the use of the SA in the present work can be justified since our main interest is to find the PM-SG phase boundary\cite{rotores,ritort}.
The
FISG model 
in the presence of transverse field is given by the Hamiltonian
\begin{equation}
\hat{H}= -\sum_{i j}J_{i j} \hat{S_{i}^{z}}\hat{S_{j}^{z}}-2\Gamma\sum_{i}\hat{S_{i}^{x}}
 \label{ee361}
 \end{equation}
where the $J_{ij}$ coupling 
is a Gaussian random variable with 
mean zero and variance $16J^{2}$. The spin  operators in Eq. (\ref{ee361}) are defined as
$\hat{S_{i}^{z}}=\frac{1}{2}[\hat{n_{i_{\uparrow}}}-\hat{n_{i_{\downarrow}}}]$
and       $\hat{S_{i}^{x}}=\frac{1}{2}[c_{i_{\uparrow}}^{\dagger}c_{i_{\downarrow}}+c_{i_{\downarrow}}^{\dagger}c_{i_{\uparrow}}]$ 
with $\hat{n}_{i\sigma}=c_{i\sigma}^{\dagger}c_{i\sigma}$ 
($\sigma=\uparrow$,
$\downarrow$).
We use  
the procedure introduced in Ref. \cite{Albaphysicaa} to obtain the grand canonical potential.
Particularly,  
in the 
1S-RSB Parisi scheme, 
the Grand Canonical Potential is written as
\begin{equation}
\begin{split}
\beta \Omega&=\frac{(\beta J)^{2}}{2}[(m-1)q_{1}^{2}-m q_{0}^{2}+\bar{q}^{2}]-\beta\mu \\ &-\frac{1}{m}\int Dz\ln \left\lbrace \int Dv [K
(z,v)
]^{m} \right\rbrace -\ln 2
\label{grandpot}
\end{split}
\end{equation}
where 
\begin{equation}
K
(z,v)
=\cosh(\beta\mu)+\int D\xi \cosh[\sqrt{\Delta
(z,v,\xi)
}],
\end{equation}
with $\Delta
(z,v,\xi)
=[\beta h
(z,v,\xi)]^{2}
+(\beta\Gamma)^{2}$ and
\begin{equation}
 h
(z,v,\xi)
=J\sqrt{2}(\sqrt{q_{0}}z+\sqrt{q_{1}-q_{0}}v+\sqrt{\bar{q}-q_{1}}\xi),
 \label{hfield}
\end{equation}
where $Dx=dx$e$^{-x^2/2}/\sqrt{2\pi}$ ($x=z,~v$ or $\xi$).
In Eqs. (\ref{grandpot}) and (\ref{hfield}), $q_0$ and $q_1$ are the 1S-RSB order parameters and $\bar{q}=q_{\alpha\alpha}=\langle S^{z}_{\alpha}S^{z}_{\alpha}\rangle$ is the diagonal replica spin-spin correlation. The parameters $q_0$, $q_1$, $\bar{q}$, and $m$ are given by the extreme condition of the grand canonical potential [Eq. \ref{grandpot}].
The RS solution is recovered when $q_0=q_1 (\equiv q)$ and $m=0$.
	In this case, the stability analysis of the RS solution is used in order to locate the tricritical point ($T_{t_{c}},\mu_{t_{c}}$) (Ref. \onlinecite{SA})  
  as a function
 of $\Gamma$.  
Therefore, the condition for all eigenvalues of the Hessian matrix to be non-negative in the PM 
solution ($q=0$)  is 
\begin{equation}
 \begin{array}{lcc}
\bar{q} < \frac{1}{\sqrt{2}\beta J} & \mbox{for} &  T/J> T_{tc}/J \\\bar{q}>\frac{f_{\phi}(T,\Gamma,\bar{q})-
  \sqrt{f_{\phi}(T,\Gamma,\bar{q})-4/(\beta J)^{2}}}{2} & \mbox{for} &  T/J< T_{tc}/J 
  \end{array}
\label{ee00}
\end{equation}
In Eq. (\ref{ee00}), the tricritical temperature $T_{tc}$ is given by
\begin{equation}
 T_{tc}/J=\frac{1}{3}\sqrt{2}f_{T_{tc}}(T_{tc},\Gamma,\bar{q}),
\label{eetric}
\end{equation}
with $f_{\phi}(T,\Gamma,\bar{q})$ defined below,
\begin{equation}
f_{\phi}(T,\Gamma,\bar{q}) = \frac{\int D\xi\left[\eta_{\phi}\cosh\sqrt{\bar{\Delta}_{\phi}} +\kappa_{\phi} \sinh\sqrt{\bar{\Delta}_{\phi}}\right] }{\int D\xi\left((\bar{h}_{\phi}^{2}/\bar{\Delta}_{\phi})\cosh \sqrt{\bar{\Delta}_{\phi}}+(\beta\Gamma)^{2}/\bar{\Delta}^{3/2}_{\phi} \right) }
\label{eefphi}
\end{equation}
where
\begin{equation} 
\begin{split}
\eta_{\phi}= & \left\lbrace \bar{h}^{4}_{\phi}+3(\beta\Gamma)^{2}\left[1-5 (\bar{h}^{2}_{\phi}/\bar{\Delta}_{\phi})  \right]\right\rbrace /\bar{\Delta}^{2}_{\phi} \\
\kappa_{\phi}= &3(\beta J)^{2}\left[2\bar{h}^{2}_{\phi}-1+5(\bar{h}^{2}_{\phi}/\bar{\Delta}_{\phi}) \right]/\bar{\Delta}^{5/2}_{\phi},
\end{split}
\end{equation}
and
\begin{equation}
  \begin{array}{lcc}
   \bar{\Delta}_{\phi}=\bar{h}^{2}_{\phi}+(\beta\Gamma)^{2}, &  &\bar{h}_{\phi}=\beta J\sqrt{2\bar{q}}\xi.
  \end{array}
\label{eedelta}
\end{equation}
In Eq. (\ref{eetric}), $f_{T_{tc}}(T_{tc},\Gamma,\bar{q})$ is given from Eq. (\ref{eefphi}), when $\phi=T_{tc}$ and $\bar{q}=1/(\sqrt{2}\beta J)$ which results in $\bar{h}_{T_{tc}}=\sqrt{\sqrt{2}\beta J}\xi$.
Furthermore, there is no stable PM solution if
\begin{equation}
 \begin{array}{lcc}
  \mu < \mu_{at}(T,\Gamma,\bar{q}) & \mbox{for}&  T/J>T_{tc}/J, \\ 
  \mu > \mu_{-}(T,\Gamma,\bar{q}) &  \mbox{for}& T/J<T_{tc}/J.
  \end{array}
  \label{eqmu}
\end{equation}
In Eq. (\ref{eqmu}), the values of $\mu_{at}$ and $\mu_{-}$ are summarized by
\begin{equation}
\mu_{\varphi}(T,\Gamma,\bar{q})=\frac{\cosh^{-1} \int D\xi [v_{\phi}\cosh\sqrt{\bar{\Delta}_{\phi}}+u_{\phi}\sinh\sqrt{\bar{\Delta}_{\phi}}]}{\beta}, 
\label{eeeq9}
\end{equation}
where
\begin{equation}
 \begin{array}{lcc}
v_{\phi}=\bar{h}^{2}_{\phi}/(\bar{q}\bar{\Delta}_{\phi})-1, &  & u_{\phi}= (\beta \Gamma)^{2}/(\bar{q}\bar{\Delta}^{3/2}_{\phi}),
  \end{array}
\end{equation}
with $\bar{\Delta}_{\phi}$ and $\bar{h}_{\phi}$ defined in Eq. (\ref{eedelta}).  
In Eq. (\ref{eeeq9}), when $\varphi=at$, $\bar{q}=1/(\sqrt{2}\beta J)$, and when $\varphi=-$, $\bar{q}=[f_{\phi}(T,\Gamma,\bar{q})-
  \sqrt{f_{\phi}(T,\Gamma,\bar{q})-4/(\beta J)^{2}}]/2$. 

In Eq. (\ref{eqmu}), $\mu_{at}$ defines the second order transition line $T_{2f}(\mu)$ and $\mu_{-}$ gives the paramagnetic spinodal line [see Figs. 1(a)-1(h)]. The  $\mu_{T_{tc}}$ value is obtained by introducing $T_{tc}$ in equation for $\mu_{at}$.

\begin{figure*}[htb]
\includegraphics[width=12.6cm,height=9cm]{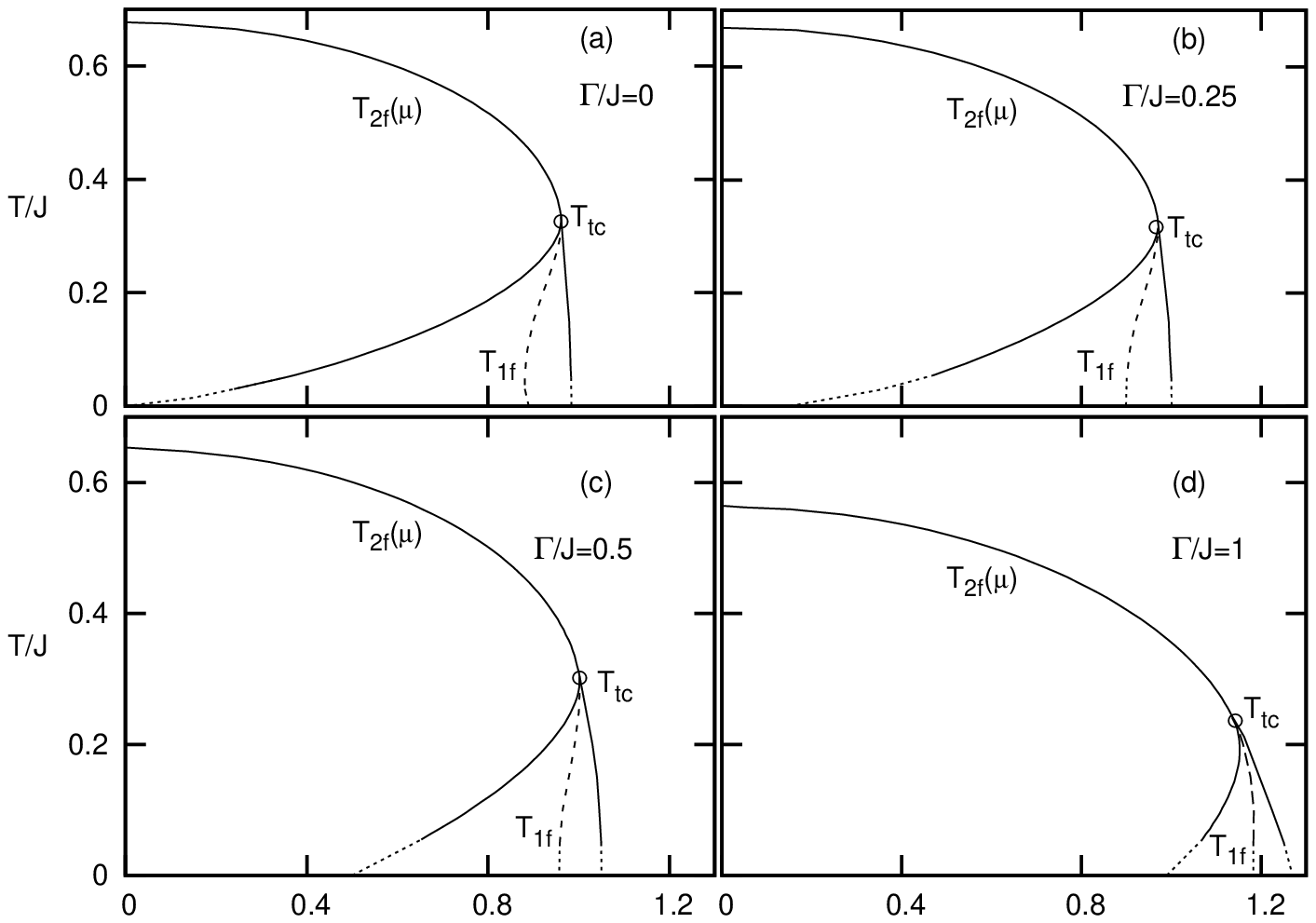}
\includegraphics[width=13cm,height=9cm]{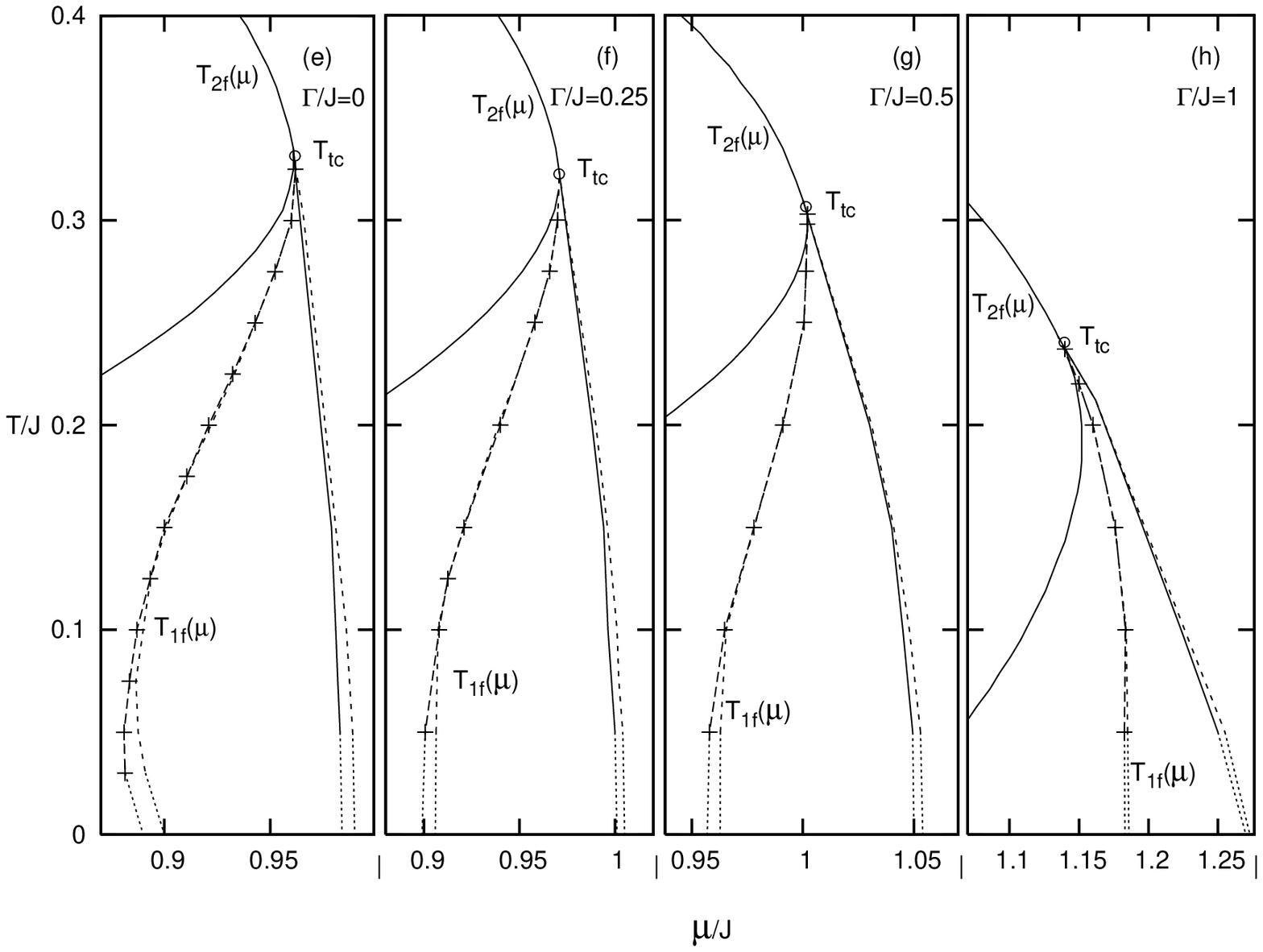}
\caption{Panels (a)-(d) show the phase diagrams $\mu/J$ versus $T/J$ for several values of  $\Gamma/J$, where $T_{2f}(\mu)$ indicates the PM/SG second order phase transition, $T_{tc}$ corresponds to the tricritical point, and $T_{1f}(\mu)$ represents the behavior of the first order transition. The spinodal lines are also exhibited (full lines below $T_{tc}$).
Panels (e)-(h) show the first order boundary 
in detail.  In these panels, $T_{1f}(\mu)$ is presented for both the RS (dashed lines) and the 1S-RSB  (pointed lines) solutions. They also exhibit the SG spinodal lines for the RS (dashed lines) and 1S-RSB  (full lines) solutions.  The dotted lines at  very low temperatures 
are extrapolations  to $T=0$.}
\label{figdiagrama21}
\end{figure*}
The PM/SG phase boundary 
in the $T/J$-$\mu/J$ plane 
within 
RS, 1S-RSB  Parisi
scheme is plotted in Figs. 1(a)-1(d)
for $\Gamma=0$, $0.25J$, $0.5J$, and $J$, respectively. 
In Figs. 1(e)-1(h) 
the position 
of 
($T_{tc}$, $\mu_{tc}$) (the tricritical point), the 
first order boundary phase and the spinodal lines are shown in detail.
In particular, when $\Gamma=0$, 
Fig. 1(a)
displays 
the first order boundary confirming the existence of a reentrance in the FISG model.
The 
turning on 
of the 
transverse field 
$\Gamma$ [see Figs. 1(b)-1(d)] produces 
a strong effect on the entire transition line. For instance,
the 
second order 
part $T_{2f}(\mu)$ 
is 
depressed.
\begin{figure}[htb]
\begin{center}
\includegraphics[width=8.6cm,height=7.5cm]{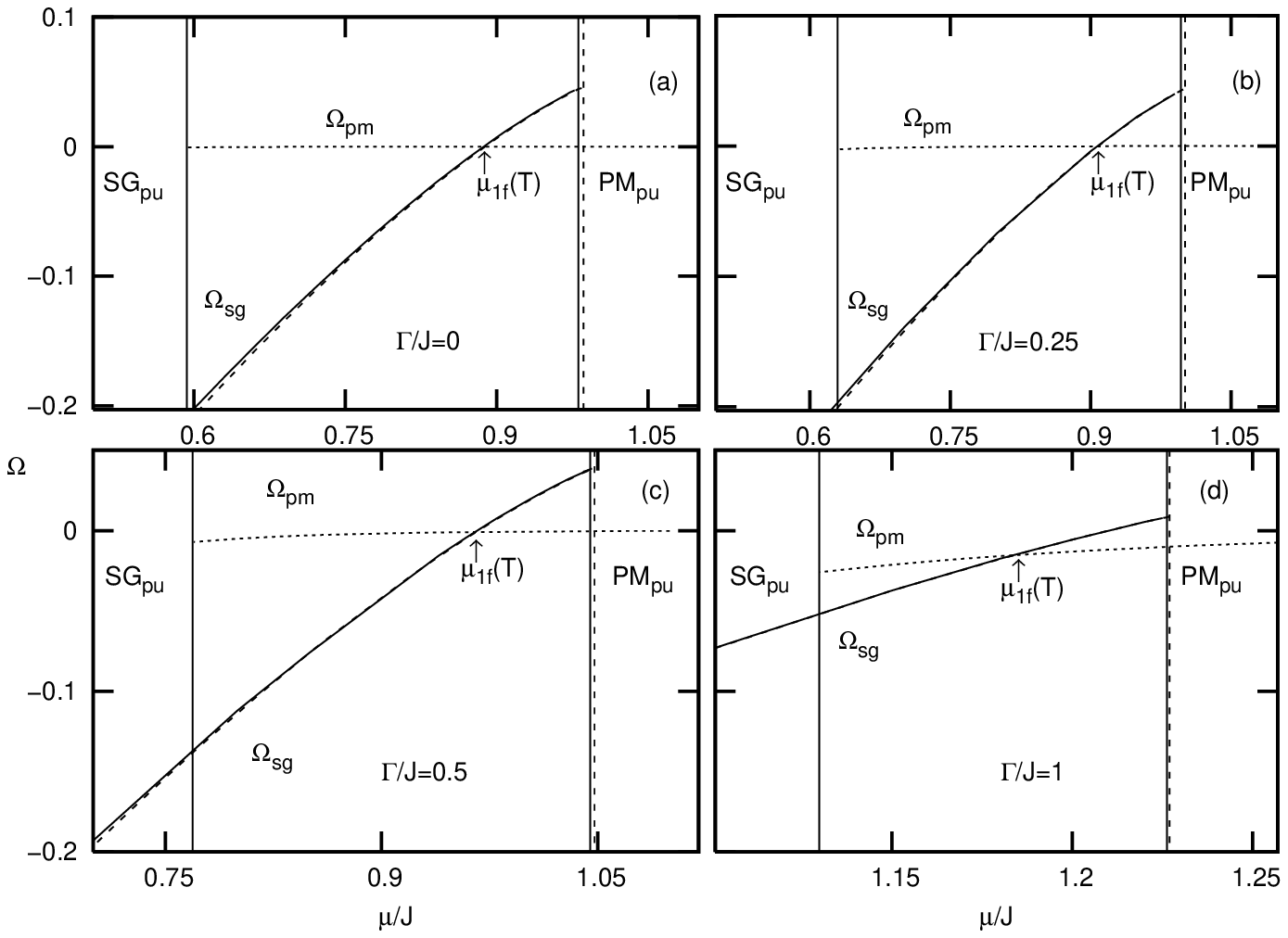}
\caption{ Grand canonical potential 
versus $\mu/J$  for several values of  $\Gamma/J$ for temperature $T/J=0.1$. 
The dotted lines represent the PM canonical potential ($\Omega_{pm}$).  The dashed and full lines represent the SG canonical potential $(\Omega_{sg})$ in the RS and 1S-RSB solutions, respectively. The vertical lines are PM (left) and SG (right) spinodal lines. The full vertical line is the 1S-RSB spinodal line.  $\mu_{1f}$ indicates the first order boundary for $T/J=0.1$. The labels $SG_{pu}$ and $PM_{pu}$  indicate the regions with only one spin-glass solution and one paramagnetic solution, respectively.}
\label{figdiagrama6}
\end{center}
\end{figure}
\begin{figure}[htb]
\begin{center}
\includegraphics[width=8.6cm,height=7.5cm]{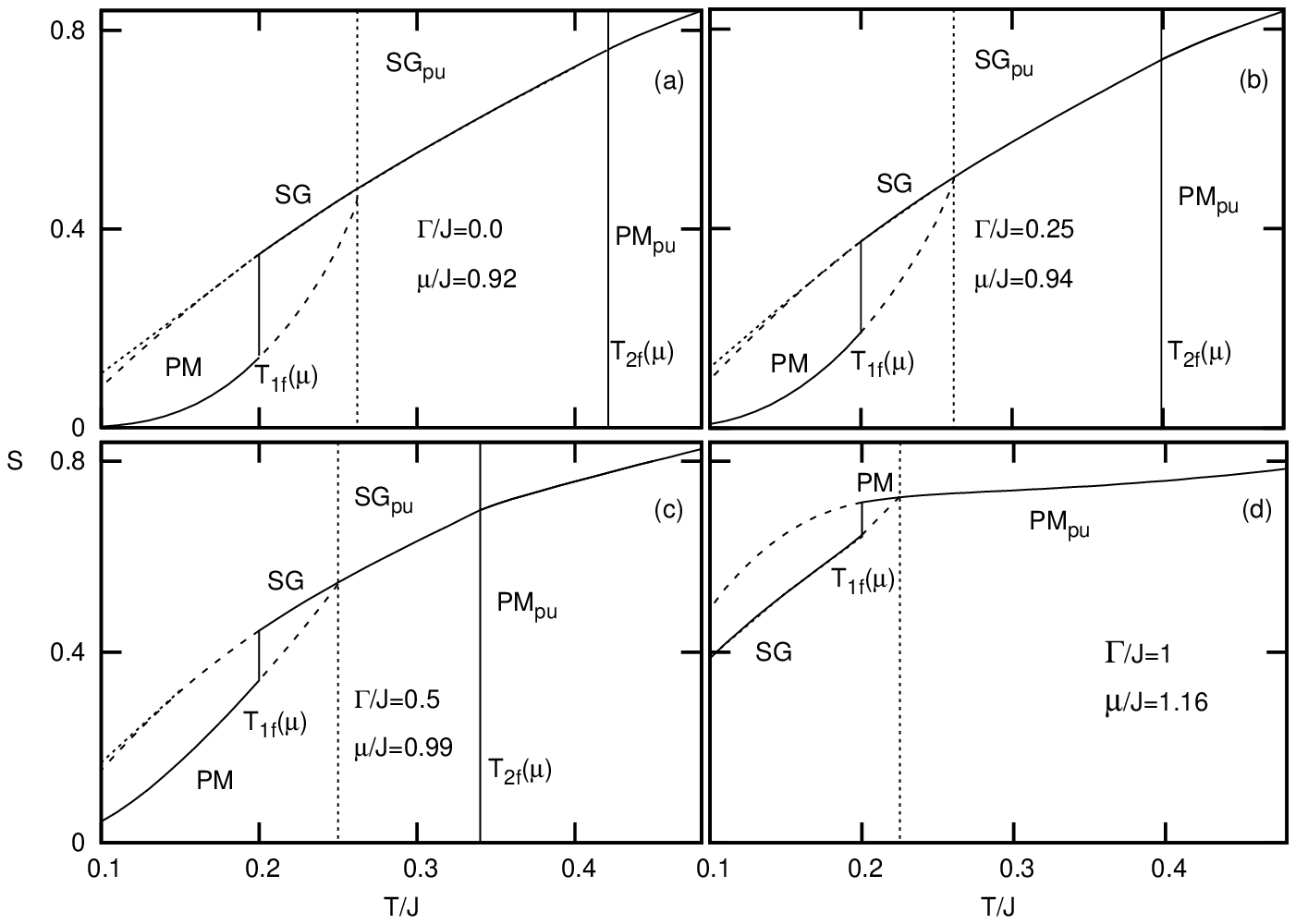}
\caption{Entropy versus $T/J$  for several values of  $\Gamma/J$ and $\mu/J$.
The first order temperature is equal to $T_{1f}(\mu)=0.2J$ for all figures.
In panels (3a)-(3c), the left vertical lines are the SG spinodal lines and the right vertical lines $T_{2f}$ are the PM/SG second order transitions. 
  At low temperatures $(T/J<0.2)$, in (3a)-(3c), the dashed lines represent the RS spin-glass solution and the dotted lines indicate the 1S-RSB spin glass solution.  
In panel (3d), the vertical line is the PM spinodal line. 
 The labels $SG_{pu}$ and $PM_{pu}$  indicate the regions with only one SG solution and one PM solution, respectively.
}
\label{figdiagrama7}
\end{center}
\end{figure}
For $\mu=0$, 
this behavior is reminiscent of the 
one 
already found 
in  
Ref. \onlinecite{Albaphysicaa} in which the freezing temperature decreases toward a QCP when $\Gamma\rightarrow \Gamma_{c}$. It should be remarked that the location of the 
tricritical point 
is 
also 
decreasing 
toward zero  while
$\Gamma$ is enhanced. 
Nevertheless, the most important 
consequence can be seen in the behavior of the first order boundary phase 
$T_{1f}(\mu)$ [or equivalently $\mu_{1f}(T)$] and,
in particular, in the reentrance  [see Figs. 1(e)-1(h)].
For $\Gamma=0.25J$ and $0.5J$, 
$T_{1f}(\mu)$
and spinodal lines are displaced in order 
to suppress the reentrance which
is completely achieved  
when
$\Gamma=J$.
In Fig. 2, the grand canonical potential 
versus $\mu/J$
is plotted at $T=0.1J$. 
 This  figure shows 
the displacement of first order boundary and 
 spinodal points which illustrate  
the gradual suppression of the reentrance exhibited in Fig. 1. In 
Figs. 1 and 2, results 
are shown 
within RS  and 1S-RSB
schemes, which  indicate that the location of the first order 
boundary is weakly dependent on the replica symmetry breaking scheme in agreement with Ref. \onlinecite{Crisanti}. There is some indication the $\Gamma$ weaken even more such dependence. However, we can not be conclusive on this point due to numerical difficulties, in particular, at low temperature. 

In Fig. 3, the entropy versus temperature is show.
The values of $\mu$ are adjusted 
to cross
the 
point $T_{1f}(\mu)=0.2J$ in the first order boundary transition.
This procedure allows us to follow the entropy difference between SG and PM phases always at the same point of the first order boundary transition.
For $\Gamma=0$, 
 we can see that
the entropy of the PM phase is found  below the SG one at the first order transition 
at  $
\mu=0.92J
$, 
which is expected from the existence of inverse freezing in the FISG  model as previously discussed.\cite{Crisanti}
The increase in $\Gamma$ produces  the decrease in the entropy difference between SG and PM phases and the displacement of the first order boundary transition 
$\mu_{1f}(0.2J)$ 
to 
larger values of $\mu$.
Simultaneously, 
the pure  PM phase (region  $PM_{pu}$ in Fig. 3) is enlarged until the total disappearance of the pure  SG phase  (region $SG_{pu}$ in Fig. 3). 
Those effects are 
related
  to the
suppression of the reentrance  shown in Figs. 1 and 2.

The previous results clearly show that the spin flipping, due to the transverse field $\Gamma$,
 suppresses
the reentrance in the first order boundary PM/SG transition and, therefore, the inverse freezing.
Actually, 
$\Gamma$ 
 depresses 
$T_{2f}(\mu)$ 
and the tricritical point
which also implies that the first order boundary appears in a 
decreasing 
interval of temperature.
It is known that, in the FISG model, 
the average occupation of the nonmagnetic states exponentially decreases with the temperature. \cite{Oppermannnuclear} 
Thus,
the enhance 
of $\Gamma$ would lead the phase transition to exist 
in
a scenario 
where the nonmagnetic states 
 are 
unimportant.  
That scenario is also consistent 
with the behavior of $n$ as a function of $\Gamma$ which has been studied  by us for several isotherms. 
In that case, 
$\Gamma$ tends to preserve the half-filling occupation even if $\mu$ increases. This effect becomes stronger when the temperature decreases. 
In that sense, 
the increase in 
$\Gamma$ 
would redistribute charge in such way that the nonmagnetic states 
become 
 gradually
avoided
at low temperature.  
There is no guarantee that the nonmagnetic states in each site have been simply excluded by 
adjusting $n$. However,  earlier results support such scenario.
In Ref. \onlinecite{Albaphysicaa}, the FISG model has been studied with an additional local
restriction
to get rid  of the nonmagnetic states. The  obtained  results 
have shown that, while 
 $\Gamma$ increases, 
the transition lines with the restriction 
and without restriction 
over the nonmagnetic sites
at the half-filling
become
increasingly close.
 
To conclude, in the present work, we have studied the role of spin flipping due to $\Gamma$ in the inverse freezing of the FISG model. It should be remarked that FISG model presents inverse freezing when $\Gamma=0$  with no need of entropic advantage.\cite{Schupper1,Crisanti}  As main result, it has been shown that $\Gamma$ destroys the reentrance in the PM/SG first order boundary and, thus, the inverse freezing.
Our results suggest that $\Gamma$ in the FISG model acts 
to redistribute
the 
charge occupation, particularly, at low temperature. In that process, the nonmagnetic states become 
unimportant for the phase transition.
In that sense, $\Gamma$ plays an opposite role concerning the inverse freezing as compared to the  degeneracy parameter 
$r$ of Ref. \onlinecite{Schupper1}. 
 Although this results are restrict to the FISG model,  we can speculate  if the suppression  of the inverse freezing by the increase in quantum effects can be a more general result.

This work was partially supported by the Brazilian agencies CNPq, CAPES, and FAPERGS.

\end{document}